\documentstyle[12pt]{article}

\begin{document}

\begin{titlepage}

\title{{\bf The electroweak phase transition with a singlet}
\thanks{Work partly supported by CICYT under
contract AEN90--0139.}}

\author{
{\bf J.R. Espinosa} \thanks{Supported by Comunidad de Madrid Grant.}
\ and \ {\bf M. Quir\'os} \\
Instituto de Estructura de la Materia. \ CSIC\\
Serrano, 123\ \ \ E--28006\ \ Madrid.\ \ Spain}

\date{}
\maketitle
\vspace{1.5cm}
\def\baselinestretch{1.15}
\begin{abstract}

We study the electroweak phase transition in the minimal
extension of the Standard Model: an extra complex singlet with
zero vacuum expectation value.  The first-order phase transition
is strengthened by the cubic term triggered in the
one-loop effective potential
by the extra boson. Plasma effects are considered to
leading order: they shield the cubic terms and weaken the
first-order phase transition. We find a region in the parameter
space where baryon asymmetry washout is avoided for Higgs
masses consistent with present experimental bounds. However in that
region the theory becomes non-perturbative for scales higher than
$10^{10}\ GeV$.

\end{abstract}

\vskip-15.cm
\rightline{{\bf hep-ph/9301285}}
\rightline{{\bf IEM--FT--67/93}}
\rightline{{\bf January 1993}}
\vskip3in

\end{titlepage}

\def\baselinestretch{1.1}

The discovery that the Standard Model
of the electroweak interactions contains an anomaly which
violates $B+L$ \cite{TH}, and that the rate of $B+L$ violation is
unsuppressed at high temperatures \cite{BL}, has revived interest
in the possibility of creating the baryon asymmetry at the
electroweak phase transition (EWPT) \cite{BAU}. In particular, the
condition that the baryon excess generated at the EWPT
is not washed out requires a strong enough first-order
phase transition,
which translates into an upper bound on the Higgs boson
mass \cite{BOUND}. Recent analyses of the minimal Standard Model (MSM)
at one-loop \cite{ONE}, and including plasma effects
\cite{T,DAISY} in various
approximations \cite{P1}-\cite{P7}, show that the above upper bound is
inconsistent with the present experimental lower bound \cite{PDG},
{\it i.e.} that the phase transition is not strongly enough
first order. Though, in our opinion, this issue is not yet fully settled
it is interesting to study extensions of the MSM where
the phase transition can be made consistent with present
experimental lower bounds on the Higgs boson mass.

In this letter we study the phase transition in the simplest
of these extensions, which consists in adding to the MSM a
complex gauge singlet
with zero vacuum expectation value. This extension was proposed
in refs.\cite{ONE,HUET} as the simplest way of overcoming the problems
encountered in the MSM. In fact the added boson generates a cubic term
in the one-loop effective potential, which can trigger a strong
first-order phase transition if it is not shielded by a
heavy $SU(2)\times U(1)$ invariant mass \footnote{This result
also holds when the gauge singlet acquires a vacuum expectation
value. This case has been recently analyzed at the one-loop
level in ref.\cite{MAJ}, where plasma effects are not considered.}.
However, as noticed in
ref.\cite{JP}, the extra boson behaves as the longitudinal
components of $W$ and $Z$ gauge bosons and
the corresponding cubic term can also be shielded
by plasma effects. Fortunately we will see this is not always
the case and find the region in parameter space where
the phase transition is strong enough first order for values of
the Higgs boson mass beyond the experimental bounds.

The lagrangian of the model is defined as:
\begin{equation}
\label{lag}
{\cal L}={\cal L}_{MSM}+\partial^{\mu}S^*\partial_{\mu}S-
M^2S^*S-\lambda_S(S^*S)^2-2\zeta^2 S^*SH^*H
\end{equation}
where $H$ is the MSM doublet with $\langle H
\rangle=\phi/\sqrt{2}$, $\phi$ is the classical field, and $M^2,
\lambda_S, \zeta^2 \ge 0$, to guarantee that $\langle S\rangle=0$ at all
temperatures \footnote{A cubic term in (\ref{lag}) would destabilize
the potential along some direction in the $S$ plane for $\lambda_S=0$.
We assume a global $U(1)$ symmetry $S\rightarrow
e^{i\alpha} S$ which prevents the appearance of such cubic term.}.

The tree-level potential is that of the MSM
\begin{equation}
\label{tree}
V_{{\rm tree}}=-\frac{\mu^2}{2}\phi^2+\frac{\lambda}{4}\phi^4
\end{equation}
and the fields contributing to the effective potential are those of
the MSM, {\it i.e.} the Higgs field $h$, the Goldstone bosons
$\vec{\chi}$, the gauge bosons $W^{\pm}, Z,\gamma$ and the top
quark $t$, with masses
\begin{equation}
\begin{array}{c}
m_h^2(\phi)=3\lambda \phi^2-\mu^2,\ \ \ m_{\chi}^2=\lambda
\phi^2-\mu^2 \\ \vspace{.5cm} {\displaystyle
m_W^2(\phi)=\frac{g^2}{4}\phi^2,\ \
m_Z^2(\phi)=\frac{g^2+g^{\prime 2}}{4} \phi^2,\ \ m_{\gamma}=0 } \\
{\displaystyle
m_t^2(\phi)=\frac{h_t^2}{2}\phi^2   },
\end{array}
\label{massMSM}
\end{equation}
(where $g,g',h_t$ are the $SU(2)\times U(1)$ gauge
and top Yukawa couplings, respectively), and the $S$ boson, with a mass
\begin{equation}
\label{massS}
m_S^2(\phi)=M^2+\zeta^2 \phi^2\ .
\end{equation}

The temperature dependent effective potential can be calculated
using standard techniques \cite{T}. Plasma effects in the leading
approximation can be accounted by the one-loop effective potential
improved by the daisy diagrams \cite{T,DAISY}. Imposing renormalization
conditions preserving the tree level values of $v^2\equiv
\mu^2/\lambda$, and working in the 't Hooft-Landau gauge, the
$\phi$-dependent part of the effective potential can be written
in the high-temperature expansion as
\begin{equation}
\label{veff}
V_{{\rm eff}}(\phi,T)=V_{{\rm tree}}+\Delta V_B+\Delta V_F
\end{equation}
where
\begin{equation}
\label{DB}
\Delta V_B= \sum_{i=h,\chi,W_L,Z_L,\gamma_L,W_T,Z_T,\gamma_T,S}
g_i \Delta V_i
\end{equation}
\begin{equation}
\label{DI}
\Delta V_i=
\left\{ \frac{m_i^2(\phi)T^2}{24}-\frac{{\cal
M}_i^3(\phi)T}{12\pi} -\frac{m_i^4(\phi)}{64\pi^2}
\left[\log\frac{m_i^2(v)}{c_BT^2}-2\frac{m_i^2(v)}{m_i^2(\phi)}
+\delta_{i\chi}\log\frac{m_h^2(v)}{m_i^2(v)} \right] \right\},
\end{equation}
where the last term comes from the infinite running of the Higgs
mass from $p^2=0$ to $p^2=m_h^2$ and cancels the logarithmic
infinity from the massless Goldstone bosons at the zero
temperature minimum, and
\begin{equation}
\label{DF}
\Delta V_F =g_t\left\{ \frac{m_t^2(\phi)T^2}{48}+\frac{m_t^4(\phi)}
{64\pi^2}
\left[\log\frac{m_t^2(v)}{c_FT^2}-2\frac{m_t^2(v)}{m_t^2(\phi)}
\right] \right\}
\end{equation}
The number of degrees of freedom $g_i$ in (\ref{DB},\ref{DF}) are
given by
\begin{equation}
\begin{array}{c}
\label{DEG}
g_h=1,\ g_{\chi}=3,\ g_S=2,\ g_t=12 \\
g_{W_L}=g_{Z_L}=g_{\gamma_L}=1, \ g_{W_T}=g_{Z_T}=g_{\gamma_T}=2
\end{array}
\end{equation}
while the coefficients $c_B$ and $c_F$ in (\ref{DI},\ref{DF})
are defined by: $\log c_B=3.9076$, $\log c_F=1.1350$.

The masses $m_i^2(\phi)$ in (\ref{DI},\ref{DF}) are defined in
(\ref{massMSM},\ref{massS}) and the Debye masses ${\cal
M}_i^2$ in (\ref{DI}) for $i=h,\chi,S,W_L,W_T,Z_T,\gamma_T$ are
\begin{equation}
\label{GAP}
{\cal M}_i^2=m_i^2(\phi)+\Pi_i(\phi,T)
\end{equation}
where the self-energies $\Pi_i(\phi,T)$ are given by
\begin{equation}
\label{PIH}
\Pi_h(\phi,T)=\left(\frac{3g^2+g^{\prime 2}}{16}+
\frac{\lambda}{2}+\frac{h_t^2}{4}+
\frac{\zeta^2}{6} \right)T^2
\end{equation}
\begin{equation}
\label{PICHI}
\Pi_{\chi}(\phi,T)=
\left(\frac{3g^2+g^{\prime 2}}{16}+
\frac{\lambda}{2}+\frac{h_t^2}{4}+
\frac{\zeta^2}{6} \right)T^2
\end{equation}
\begin{equation}
\label{PIS}
\Pi_S(\phi,T)=\frac{\lambda_S+\zeta^2}{3}T^2
\end{equation}
\begin{equation}
\label{PIWL}
\Pi_{W_L}(\phi,T)=\frac{11}{6}g^2 T^2
\end{equation}
\begin{equation}
\label{PIT}
\Pi_{W_T}(\phi,T)=\Pi_{Z_T}(\phi,T)=\Pi_{\gamma_T}(\phi,T)=0
\end{equation}

The Debye masses ${\cal M}_i^2$ for $i=Z_L,\gamma_L$ are given by
\begin{equation}
\label{GAPL}
\left(
\begin{array}{cc}
{\cal M}^2_{Z_L} & 0 \\
0 & {\cal M}^2_{\gamma_L}
\end{array}
\right)=R(\theta_W^{(1)})
\left(
\begin{array}{cc}
m_Z^2(\phi)+\Pi_{Z_L Z_L} & \Pi_{Z_L\gamma_L} \\
\Pi_{\gamma_L Z_L} & \Pi_{\gamma_L \gamma_L}
\end{array}
\right)
R^{-1}(\theta_W^{(1)})
\end{equation}
with the rotation $R(\theta_W^{(1)})$
\begin{equation}
\label{ROT}
R(\theta_W^{(1)})=
\left(
\begin{array}{cc}
\cos \theta_L^{(1)} & -\sin \theta_L^{(1)} \\
\sin \theta_L^{(1)} & \cos \theta_L^{(1)}
\end{array}
\right)
\end{equation}
and the self-energies \footnote{We correct the coefficients of the
last terms in (\ref{PIZZ}), $\frac{8}{3}$, and in (\ref{PIGG}),
$\frac{11}{3}$, which were misprinted
as $4$ and $\frac{15}{3}$, respectively, in \cite{P3}.}
\begin{equation}
\label{PIZZ}
\begin{array}{ll}
\Pi_{Z_L Z_L}(\phi,T)
 &
=\left({\displaystyle \frac{2}{3}g^2\cos^2\theta_W
+\frac{1}{6}\frac{g^2}{\cos^2 \theta_W}(1-2\sin^2 \theta_W
\cos^2 \theta_W) }
  \right. \\
 &\left.
{\displaystyle
+\frac{g^2}{\cos^2 \theta_W}
(1-2\sin^2 \theta_W+\frac{8}{3}\sin^4 \theta_W) } \right)T^2
\end{array}
\end{equation}
\begin{equation}
\label{PIGG}
\Pi_{\gamma_L \gamma_L}(\phi,T)=\frac{11}{3}e^2 T^2
\end{equation}
\begin{equation}
\label{PIGZ}
\Pi_{\gamma_L Z_L}(\phi,T)=\frac{11}{6}eg
\frac{\cos^2 \theta_W-\sin^2 \theta_W}{\cos\ \theta_W}T^2
\end{equation}
The angle $\theta_L^{(1)}$ in (\ref{GAPL}) is the one-loop
temperature dependent correction to the electroweak angle. In
fact the angle $\theta_L(\phi,T)$ defined by
\begin{equation}
\label{TANGLE}
\theta_L(\phi,T)=\theta_W+\theta_L^{(1)}
\end{equation}
maps $(A_3,B)$ into $(Z,\gamma)$.

Using (\ref{PIZZ}-\ref{PIGZ}) one obtains the eigenvalues and
rotation angle in (\ref{GAP}) as:
\begin{equation}
\label{MZL}
{\cal M}_{Z_L}^2=\frac{1}{2}\left[m_Z^2(\phi)+
\frac{11}{6}\frac{g^2}{\cos^2 \theta_W}T^2+\Delta(\phi,T)
\right]
\end{equation}
\begin{equation}
\label{MGL}
{\cal M}_{\gamma_{L}}^2=\frac{1}{2}\left[m_Z^2(\phi)+
\frac{11}{6}\frac{g^2}{\cos^2 \theta_W}T^2-\Delta(\phi,T)
\right]
\end{equation}
\begin{equation}
\label{ANGLE}
\sin\ 2\theta_L^{(1)}(\phi,T)=-\frac{2\Pi_{\gamma_L Z_L}}{\Delta}
\end{equation}
\begin{equation}
\label{SIN}
\sin\ 2\theta_L(\phi,T)=\sin\ 2\theta_W
\frac{m_Z^2(\phi)}{\Delta(\phi,T)}
\end{equation}
with
\begin{equation}
\label{DELTA}
\Delta^2(\phi,T)=m_Z^4(\phi)+\frac{11}{3}\frac{g^2\cos^2
2\theta_W}{\cos^2 \theta_W} \left[m_Z^2(\phi)+\frac{11}{12}
\frac{g^2}{\cos^2 \theta_W} T^2 \right]T^2
\end{equation}
It is clear from (\ref{SIN},\ref{DELTA}) that at zero
temperature the electroweak angle coincides with the usual one:
$\Delta(\phi,0)=m_Z^2(\phi)$,
$\theta_L(\phi,0)\equiv \theta_W$.

An analytic treatment of the one-loop effective potential was
given in ref.\cite{ONE}. In the presence of plasma effects a
similar treatment of the potential is not available.
Before performing the complete
numerical analysis it is instructive to get an
analytic feeling of the efectiveness of the screening provided
by plasma effects. This can be done assuming that the
bosonic contribution (\ref{DB}) to the effective
potential (\ref{veff}) is dominated by one field, namely the $S$
field, and neglecting the contribution from the other bosons.
The result can be used to illustrate other physical situations where the
effective potential is dominated by one kind
of bosonic fields \footnote{A good example is the case of
the minimal supersymmetric standard model \cite{MSSM} where the
bosonic part of the effective potential can be dominated by the
contribution of squarks.}.
The $\phi$ dependent part of the
effective potential (\ref{veff}) can be written as
\begin{equation}
\label{veffapp}
V(\phi)=A(T)\phi^2+B(T)\phi^4+C(T)\left(\phi^2+K^2(T)\right)^{3/2}
\end{equation}
where
\begin{equation}
\label{A}
A(T)=-\frac{1}{2}\mu_T^2+\frac{1}{4}\left(\frac{\zeta^2}{3}
+\frac{h_t^2}{2}\right)T^2
\end{equation}
\begin{equation}
\label{B}
B(T)=\frac{1}{4}\lambda_T
\end{equation}
\begin{equation}
\label{C}
C(T)=-\frac{\zeta^3T}{6\pi}
\end{equation}
\begin{equation}
\label{K2}
K^2(T)=\frac{(\zeta^2+\lambda_S)T^2+3M^2}{3\zeta^2}
\end{equation}
and
\begin{equation}
\label{MUT}
\mu_T^2=\mu^2-\frac{\zeta^2}{8\pi^2}\left\{m_S^2(v)+M^2\log\frac{c_BT^2}
{m_S^2(v)}\right\}+\frac{3}{8\pi^2}h_t^2m_t^2(v)\log\frac{m_t^2(v)}{c_FT^2}
\end{equation}
\begin{equation}
\label{LAMBDAT}
\lambda_T=\lambda+\frac{\zeta^4}{8\pi^2}\log\frac{c_BT^2}{m_S^2(v)}
+\frac{3}{16\pi^2}h_t^4\log\frac{m_t^2(v)}{c_FT^2}
\end{equation}

The temperature $T_2$ is defined by the condition $V''(0)=0$, or
\begin{equation}
\label{T2}
4A^2-9C^2K^2=0
\end{equation}
For $T<T_2$ the origin is a maximum, and there is a global
minimum at $\phi \neq 0$ that evolves towards the zero
temperature minimum. For $T>T_2$ the origin is a minimum and
there is a maximum at $\phi_-(T)$ and a minimum at $\phi_+(T)$ given by
\begin{equation}
\label{mM}
\phi_{\pm}^2(T)=\frac{1}{32B^2}\left\{9C^2-16AB\pm 3|C|
\sqrt{9C^2+32(2B^2K^2-AB)} \right\}
\end{equation}
At the temperature $T_1$ defined by the condition
\begin{equation}
\label{T1}
9C^2+32(2B^2K^2-AB)=0
\end{equation}
the maximum and minimum collapse $\phi_-(T_1)=\phi_+(T_1)$. For
$T>T_1$ the origin is the only minimum.

Using (\ref{A}-\ref{K2}) the temperatures $T_1$ and $T_2$ can be
written as
\begin{equation}
\label{TT1}
\zeta^2T_1^2=\frac{2\lambda_{T_1}(\zeta^2\mu_{T_1}^2+\lambda_{T_1}M^2)}
{(\frac{\zeta^2}{3}+\frac{h_t^2}{2})\lambda_{T_1}
-\frac{\zeta^6}{8\pi^2}-\frac{2
\lambda_{T_1}^2}{3\zeta^2}(\zeta^2+\lambda_S)}
\end{equation}
\begin{equation}
\label{TT2}
T_2^2=\frac{1}{2\alpha}\left\{\Lambda^2(T_2)+\sqrt{\Lambda^4(T_2)
-16\alpha \mu^4_{T_2}}\right\}
\end{equation}
where
\begin{equation}
\label{ALPHA}
\alpha=\left(\frac{\zeta^2}{3}+\frac{h_t^2}{2}\right)^2-
\frac{1}{3\pi^2}\zeta^4\left(\zeta^2+\lambda_S\right)
\end{equation}
\begin{equation}
\label{LAMBDA}
\Lambda^2(T)=\frac{1}{\pi^2}\zeta^4M^2+4\left(\frac{\zeta^2}{3}+
\frac{h_t^2}{2}\right)\mu_T^2
\end{equation}

The nature of the phase transition depends on the relation
between $T_1$ and $T_2$. For values of the parameters
$(\zeta^2,\lambda_S,M)$ such that $T_1>T_2$ the phase transition
is first order and the plasma screening is not very effective.
When $T_1=T_2$ the phase transition becomes second order because the
screening became more effective. In fact, the condition $T_1=T_2$ gives the
turn-over from first to second order. It provides a surface in the
space $(\zeta,\lambda_S,M)$ which separates first-order from
second-order regions. The different regions in
the $(\zeta,\lambda_S)$-plane are plotted in Fig.1 for
$m_h=65\ GeV$, $m_t=120\ GeV$ and different values of $M$. An
analytic approximation can be given, if one neglects loop corrections
in (32,33), as
\begin{equation}
\frac{\zeta^8}{8\pi^2}-\frac{2}{3}\lambda^2(\zeta^2+\lambda_S)\geq
\left[
\lambda\left(\frac{\zeta^2}{3}+\frac{h_t^2}{2}\right)
-\frac{\zeta^6}{4\pi^2} \right]
\left(\frac{M}{v}\right)^2
\end{equation}
where the strict inequality corresponds to the subregion of the space
$(\zeta,\lambda_S,M)$ for which the phase transition is first-order,
and the equality corresponds to the turn-over to a second order phase
transition.
Since the right-hand side of eq. (41) is
positive-definite \footnote{It is easy to see that the necessary and
sufficient condition for this turn-over to exist is that the
right-hand-side of eq. (41) is positive-definite. It is not a priori
excluded that an isolated region exists in the space
$(\zeta,\lambda_S,M)$ where the phase transition is second order.
However we have checked that for phenomenological values of the
parameters $(\lambda,h_t)$ this region does not exist.}, we can see
from (41) how the parameters
$M^2$ and $\zeta^2+\lambda_S$ in (4) and (5) influence the shielding
of the first order phase transition. First, the larger the value of
$M$ is, the easier one saturates the inequality in (41) and the
easier one reaches a second-order phase transition. The same can be
stated on $\lambda_S$, though its effect is damped by $\lambda^2$ and
would become important only for a very heavy Higgs. For the same
reason the effect of $\zeta$ is opposite, unless the Higgs is very
heavy. These effects can be read off from Fig. 1 where no
approximations are introduced.

The complete numerical analysis of eq.(\ref{veff}) is summarized
in Figs.2-5 \footnote{We have used, in Figs.2-5, the numerical computation
of the integrals \cite{T} giving rise to the high-temperature expansion
of eqs.(\ref{DI},\ref{DF}).}
where we plot $\phi_{+}(T_c)/T_c$, where $T_c$ is
the temperature at which the minima at $\phi_+$ and at the
origin are degenerate, for different regions of the space of
parameters. In Figs.2 and 3  we plot $\phi_{+}(T_c)/T_c$ versus
$\zeta^2$ for $M=50\ GeV$, $0\le\lambda_S \le 1$ and
$\lambda_S=0$, $0 \le M\le 1\ TeV$, respectively, and
$m_h=65\ GeV$, $m_t=120\ GeV$. We can check from Fig.3 that for values
of $M$ much greater than the electroweak scale, the $S$ field decouples and
one recovers the MSM result.

For our choice of lagrangian parameters in (\ref{lag}), in
particular for $M^2>0$, the sphaleron energy is
minimized for field configurations with $S\equiv 0$ \cite{KZ},
which are the usual sphalerons in the MSM \cite{BL}. In that
case the condition $E_{{\rm sph}}(T_c)/T_c>45$ \cite{BOUND}
corresponds to
$\phi_{+}(T_c)/T_c>(\phi_{+}(T_c)/T_c)_{{\rm min}}$, which is
plotted in Fig.4 (dashed line) versus $m_h$.

If we want to establish an absolute upper bound on the
mass of the Higgs boson we need to optimize the phase transition
with respect to the new parameters $(\zeta,\lambda_S,M)$. As can
be seen from Figs.2 and 3 this is accomplished for $M=0$
\footnote{We are aware that values $M\ll v$ would require much
more fine tuning than that required for the Higgs sector of the
MSM. However we are taking $M=0$ only to establish an absolute
upper limit on the Higgs mass.} and $\lambda_S =0$.
As for $\zeta$, a quick glance at Figs.2 and 3 shows that we
should put it to its maximum value $\zeta_{{\rm max}}$. The
usual requirement for $\zeta_{{\rm max}}$ is that the theory
remains perturbative in all its domain of validity, from the
electroweak scale to a high scale $\Lambda$. For that we have to
study the renormalization group equations (RGE) of the minimal
extension of the MSM provided by the lagrangian (\ref{lag}).
At one-loop the only $\beta$-function of the MSM modified by the
interactions of $S$ is $\beta_{\lambda}$ while there appear new
$\beta$-functions for the new couplings $\lambda_S$ and $\zeta$,
as \cite{CEL}
\begin{equation}
\label{bl}
\Delta \beta_{\lambda}=8\zeta^4
\end{equation}
\begin{equation}
\label{bls}
\beta_{\lambda_S}=20\lambda_S^2+8\zeta^4
\end{equation}
\begin{equation}
\label{bz}
\beta_{\zeta^2}=\zeta^2\left[6\lambda+6h_t^2+8\lambda_S+8\zeta^2-
\frac{3}{2}(3g^3+g'^2)\right]
\end{equation}
where we are using the convention
\begin{equation}
\label{RGE}
16\pi^2 \frac{dx}{dt}=\beta_x
\end{equation}
for all couplings $x=\zeta^2,\lambda_S,\lambda,...$.
{}From eqs.(\ref{bls},\ref{bz}) we see that imposing
$\lambda_S(M_W)=0$ as boundary condition, consistent with our
previous requirement, we can reach the maximum value of
$\zeta(M_W)$, $\zeta_{{\rm max}}$, that will depend on $m_h$,
$m_t$ and $\Lambda$. We have solved the system of RGE
corresponding to the lagrangian (\ref{lag}) between $M_W$ and
$\Lambda$ and obtained $\zeta_{{\rm max}}$ for different values
of $m_h$ and $m_t$. The dependence of $\zeta_{{\rm max}}$ on
$m_h$ is negligible for $60\ GeV \le m_h \le 100\ GeV$. In
Table 1 we show $\zeta^2_{{\rm max}}$ for different values of
$\Lambda$ and $m_t=90,\ 120,\ 175\ GeV$.
In Fig.4 we plot
$\phi_{+}(T_c)/T_c$ versus $m_h$ for $\lambda_S=0$, $\zeta=
\zeta_{{\rm max}}$, as taken from Table 1 for the different values
of $\Lambda$, $M=0$ and $m_t=120\ GeV$.
To exhibit the dependence on $m_t$ we plot in Fig.5
$\phi_{+}(T_c)/T_c$ versus $m_h$ for $\lambda_S=0$, $M=0$,
$m_t=90,\ 120$ and 175 $GeV$, and $\zeta=
\zeta_{{\rm max}}$, corresponding to $\Lambda=10^{6}\ GeV$. In that case
we see from Fig.5 that avoiding baryon asymmetry washout imposes on the
Higgs mass an upper bound of order $80\ GeV$.

In conclusion, we have obtained that the minimal extension of the
MSM we have analyzed in this paper is consistent with $\Delta(B+L)$
and the experimental bounds on the Higgs mass provided the theory
remains valid up to a scale $\Lambda \le 10^{10}\ GeV$. In the case the
theory remains valid up to a higher scale (as {\em e.g.} the Planck scale)
the former conditions would force some coupling constants to become
non-perturbative below the high scale. Generically this would require extra
physics between $\Lambda$ and the Planck scale.

\newpage
\section*{Table captions}
\begin{description}
\item[Table 1]
$\zeta^2_{{\rm max}}$ for different values of $\Lambda$ and $m_t$.
\end{description}
\section*{Figure captions}
\begin{description}
\item[Fig.1]
Plots of the condition $T_1=T_2$, eqs.(\ref{TT1},\ref{TT2}), in
the $(\zeta,\lambda_S)$ plane for $m_h=65\ GeV$, $m_t=120\ GeV$
and values of $M$ from 0 to $300\ GeV$, with a step
of $50\ GeV$. The lower (upper) curve
corresponds to $M=0$ ($M=300\ GeV$).
\item[Fig.2]
Plot of $\phi_{+}(T_c)/T_c$ as a function of $\zeta^2$ for
$\lambda_S=0$ (upper curve), $0.5$ and $1.0$ (lower curve),
$M=50\ GeV$, $m_h=65\ GeV$ and $m_t=120\ GeV$.
\item[Fig.3]
The same as in Fig.2, but for $\lambda_S=0$,
$m_h=65\ GeV$, $m_t=120\ GeV$ and $M=0$ (upper curve), 50, 100,
150, 200, 250, 300, 350, 400 and 1000 $GeV$ (lower curve).
\item[Fig.4]
Plot of $\phi_{+}(T_c)/T_c$ versus $m_h$ for $\lambda_S=0$, $M=0$,
$m_t=120\ GeV$ and $\zeta=
\zeta_{{\rm max}}$, as taken from Table 1 for  $\Lambda=
10^4,\ 10^6,\ 10^8,\ 10^{10},\ 10^{12},\ 10^{14}$ and $10^{16}\ GeV$.
\item[Fig.5]
The same as in Fig.4, but for $m_t=90,\ 150$ and $175\ GeV$,
$\Lambda= 10^6\ GeV$ and $\zeta_{{\rm max}}$ taken from Table 1.
\end{description}
\newpage
\begin{center}
\begin{tabular}{|c|c|c|c|} \hline
                  &  \multicolumn{3}{c|}{$m_t\ (GeV)$}\\ \cline{2-4}
$\Lambda\ (GeV)$  &    90       &     120         &     175   \\
\hline\hline
$10^4$    & 1.774 & 1.742 & 1.667 \\
$10^6$    & 1.095 & 1.067 & 1.011 \\
$10^8$    & 0.793 & 0.770 & 0.728 \\
$10^{10}$ & 0.624 & 0.604 & 0.573 \\
$10^{12}$ & 0.515 & 0.498 & 0.473 \\
$10^{14}$ & 0.439 & 0.424 & 0.405 \\
$10^{16}$ & 0.384 & 0.370 & 0.356 \\ \hline
\end{tabular}
\end{center}
\vspace{1cm}
\begin{center}
\Large
{\bf Table 1}
\end{center}
\end{document}